\documentstyle[prl,aps]{revtex}

\begin{document}
\draft
\twocolumn[
\widetext

\title{Ginzburg-Landau Theory of Vortices in $d$-wave
Superconductors}
\author{A. J. Berlinsky$^1$, A. L. Fetter$^2$, M. Franz$^1$, C. Kallin$^1$,
and P. I. Soininen$^1$}
\address{$^1$Institute for Materials Research and Department of
Physics and Astronomy,
McMaster~University,~Hamilton,~Ontario,~L8S~4M1~Canada}
\address{$^2$Department of Physics, Stanford University, Stanford, CA
94305}

\date{\today}
\maketitle

\widetext
\begin{abstract}
\leftskip 54.8pt
\rightskip 54.8pt

Ginzburg-Landau  theory is used to study the properties of single
vortices and of the Abrikosov vortex lattice in a $d_{x^2-y^2}$
superconductor.  For a single vortex, the $s$-wave order parameter has the
expected four-lobe structure in a ring around the core and falls
off like $1/r^2$ at large distances.
The topological structure of the
$s$-wave order parameter consists of one counter-rotating unit vortex,
centered at the core, surrounded by four symmetrically placed positive
unit vortices. The
Abrikosov  lattice is shown to have a triangular structure close to
$T_c$ and an oblique structure at lower temperatures. Comparison
is made to recent neutron scattering data.
\end{abstract}


]

\widetext
\begin{minipage}[t]{3.4in}
\setlength{\parindent}{5pt}
\leftskip -10pt
\rightskip 10pt
Evidence continues to accumulate for the existence of nodes in
the gap of high $T_c$ superconductors\cite{nodes}, although the
nature of the microscopic mechanism remains controversial.  Thus
there is good reason to study the phenomenology of
superconductors with non-trivial gap structures, particularly
their behavior in an applied magnetic field that penetrates the
superconductor, creating vortices which then play a dominant role
in the transport properties of the system. We have previously
considered\cite{soininen} a simple microscopic model of $d$-wave
superconductivity for electrons on a lattice and used the
Bogoliubov-de Gennes equations to calculate the order parameter
distribution for a single vortex. The relevant Ginzburg-Landau
(GL) free energy, which was first derived by Joynt\cite{joynt},
served to interpret our results. The GL theory involves both the
$d$-wave order parameter and an induced $s$-wave order parameter
which arises through a mixed gradient coupling.

\vspace{-4pt}
\noindent\hrulefill
\end{minipage} \
\begin{minipage}[t]{3.4in}
\setlength{\parindent}{5pt}
Experience with conventional ($s$-wave) superconductors
has demonstrated that virtually all of the
phenomenological properties  of superconductors can be derived
from the appropriate
GL theory.  Here we present analogous results for what is arguably the
simplest model of an unconventional superconductor with a
non-trivial gap structure.  First we integrate the GL equations for a
single vortex numerically.  The results, shown in Figure 1, exhibit
a non-trivial internal topological structure for the $s$-wave
component. Second we solve for the structure of the vortex lattice
close to $H_{c2}(T)$, by first minimizing the quadratic part of the
free energy using a simple variational wave function and then
forming a periodic array of vortices from linear combinations of
these functions.  The results are compared to recent small angle
neutron scattering (SANS) data \cite{keimer1}.

The free energy of a $d_{x^2-y^2}$ superconductor may be expressed
in terms of the two order parameters, $s({\bf r})$ and
$d({\bf r})$, with appropriate symmetries, as follows \cite{joynt}
\end{minipage}
\widetext
\begin{eqnarray}
f= \alpha_s|s|^2 &+& \alpha_d|d|^2 +\beta_1|s|^4 + \beta_2|d|^4 +
     \beta_3|s|^2|d|^2 +\beta_4(s^{*2}d^2 + d^{*2}s^2) \nonumber \\
  &+&\gamma_s|{\vec\Pi} s|^2 + \gamma_d|{\vec\Pi} d|^2
     +\gamma_v\bigl[ (\Pi_y s)^*(\Pi_y d) - (\Pi_x s)^*(\Pi_x d) + {\rm c.c.}
\bigr].
\label{fgl}
\end{eqnarray}
Here ${\vec\Pi}=-i\hbar\nabla -e^*{\bf A}/c$, and
$d$ is assumed to be the critical order parameter, i.e., we take
$\alpha_s=T-T_s$, $\alpha_d =
T-T_d$ with
 $T_s<T_d$. It is also assumed that $\beta_1$, $\beta_2$, $\beta_3$,
$\gamma_s$, $\gamma_d$ and $\gamma_v$ are all
positive\cite{soininen}.  The
parameters
$\gamma_i$ are related to the effective masses in the usual way. We use
$\gamma_i=\hbar^2/2m_i^*$, for
$i=s,d,v$.  The field equations
for the order parameters are obtained by varying the
free energy (\ref{fgl}) with respect to conjugate fields $d^*$ and
$s^*$, giving,
\begin{mathletters}
\label{eq:sd}
\begin{equation}
\biggl({\hbar^2\over 2m_d^*} \Pi^2 + \alpha_d\biggr) d + {\hbar^2\over 2m_v^*}
(\Pi_y^2-\Pi_x^2)s + 2\beta_2|d|^2d + \beta_3|s|^2d + 2\beta_4s^2d^* = 0,
\label{eq:d}
\end{equation}
\begin{equation}
\biggl({\hbar^2\over 2m_s^*} \Pi^2 + \alpha_s\biggr)s + {\hbar^2\over 2m_v^*}
(\Pi_y^2-\Pi_x^2)d +2\beta_1|s^2|s + \beta_3|d|^2s + 2\beta_4d^2s^* = 0.
\label{eq:s}
\end{equation}
\end{mathletters}
\noindent Equations (\ref{eq:sd})  can be integrated
numerically for boundary conditions which generate a single vortex
at the origin. In doing this, we assume an extreme type-II limit,
where the coupling to the vector potential can be ignored while
considering the core structure of the isolated vortex line.
\twocolumn[]
\narrowtext

Ren {\em et al.} \cite{Ren} have previously shown that
for a $d$-wave order parameter with the asymptotic form,
\begin{equation}
d(r,\theta)=d_0 e^{i \theta},
\end{equation}
where $d_0=\sqrt {-\alpha_d /2 \beta _2} $,
the asymptotic form of the $s$-wave order parameter is:

\begin{equation}
s(r,\theta) = g_1(r)e^{-i \theta}+g_2(r)e^{i 3 \theta},
\end{equation}

\noindent where $g_1(r)$ and $g_2(r)$ fall off like $1/r^2$ for large $r$.
Furthermore, close to $T_d$, $g_2(r) \approx -3 g_1(r)$ and
therefore the winding number far from the core is $+3$.
This result combined with the result that close to the core the
winding number is $-1$\cite{Volovik} forces us to conclude that four
additional positive vortices must exist outside the core. We
emphasize that this is a topological result and thus not sensitive to
small modifications of the parameters.
As is shown below, these
vortices lie on the $\pm x$ and $\pm y$ axes.
At lower temperatures
a topological transition
to a state with $s$-wave
winding number
$-1$ is in principle possible.

Asymptotically the
superconductor is not in a pure $d$-wave state, but rather in a state
characterized by power law decay of the $s$-wave component.
Only at  the length scale given by the penetration depth is the pure $d$-wave
state regained.

We have studied the dependence of the maximum of the $s$-wave
component on the GL parameters. Noting that both the $d$-wave and
$s$-wave components rise over the same length scale given by $\xi_d$,
where $\xi_d^2 = \gamma_d / |\alpha _d|$, allows us to give an order
of magnitude estimate for the magnitude of the $s$-wave order
parameter at the maximum,
\begin{equation}
{\max(s)\over d_0} \sim {\gamma_v \over \alpha_s \xi_d^2}.
\end{equation}
Our numerical results confirm that the constant of
proportionality is of the order unity. Note that the temperature
dependence of $\max(s)$ is  $(1-T/T_d)^{3/2}$.

In Figure \ref{fig1} we show the behavior of the $s$-wave
amplitude along the $x$-axis
and along the diagonal, as obtained by numerical integration of
Eqs.(\ref{eq:sd}).
 Moving outward from the center of the vortex
both the $s$- and $d$-wave amplitudes increase over the same length
scale $\xi_d$. Moving further out one enters a region where the
relative phase tends to lock to value $\pm \pi/2$. The change in the
relative phase takes place in narrow ``domain walls''. Up to this
distance the results are in perfect agreement with the ones obtained
within Bogoliubov-de Gennes theory\cite{soininen}. However, further
out the situation changes. The domains of rapid variation of the
relative phase vanish. Furthermore, the relative phase starts to
wind in the opposite direction. In Figure  \ref{fig1} this change manifests
itself as a zero in the amplitude of the $s$-wave component. This zero
is nothing but the core of one of 4 ``extra'' vortices.
Identical vortices are found at all the four ``wall ends''; this
combined with the vortex with an opposite charge at the center
gives the total required winding of $+3$.

Next we turn to the problem of the structure of the vortex lattice
in the vicinity of the upper critical field $H_{c2}$ where the
amplitudes of the order parameters are small and it is sufficient to
consider the linearized GL equations. It is easily seen
that in the Landau gauge (${\bf A}={\hat y} Bx$)
these linearized field  equations are satisfied by taking
$d({\bf r})=e^{iky}d(x),\ \
s({\bf r})=e^{iky}s(x)$. Then, exactly as in the one component case
\cite{abrikosov}, we
are left with a one dimensional problem which can be stated as
follows:

\begin{mathletters}
\label{eq:sdlin2}
\begin{equation}
({\cal H}_0+\alpha_d) d + Vs = E d,
\label{eq:sdlin2:d}
\end{equation}
\begin{equation}
Vd + ({\cal H}_0+\alpha_s)s = E s,
\label{eq:sdlin2:s}
\end{equation}
\end{mathletters}
\noindent where ${\cal H}_0 = \hbar\omega_c(a^\dagger a +1/2)$ and
$V={\epsilon_v}(\hbar\omega_c/2) (a^\dagger a^\dagger +aa)$ are expressed in
terms of the usual raising and lowering operators, which can be
written as $a=[(x-x_k)/l + l(\partial/\partial x)]/\sqrt{2}$. Here
$l=\sqrt{\hbar c/e^* B}$ is the magnetic length, $x_k=kl^2$ and
$\omega_c=(e^*B/mc)$. In writing Eqs. (\ref{eq:sdlin2}), we have
assumed, for simplicity, that $m^*_d=m^*_s \equiv m$ , i.e. that
$\gamma_d=\gamma_s$, and we have set ${\epsilon_v} = \gamma_v/\gamma_s =
m^*_s/m^*_v$.
By including the right hand side  of Eqs.(\ref{eq:sdlin2})
we  are considering a slightly more general problem: $E=0$ corresponds to the
physical solution for $B=H_{c2}(T)$, and solutions for $E<0$ will be
useful later when we consider the
stability of various vortex lattice structures.

In contrast to the one component case, the linearized Eqs.(\ref{eq:sdlin2})
have no obvious exact solutions. This is due to the coupling term $V$ whose
origin traces back to the mixed gradient term in the free energy (\ref{fgl}).
In what follows we construct a simple variational solution, which is likely to
capture all the essential physics of the problem. To this end we define
${\cal H}^\pm = {\cal H}_0 \pm V$, and $ \varphi^\pm = d \pm s$,
in terms of which we can write the set of equations (\ref{eq:sdlin2}) as
\begin{displaymath}
\left( \begin{array}{cc}
{\cal H}^++T-T^*  & -{\Delta T}/2 \\
                          &      \\
-{\Delta T}/2              & {\cal H}^-+T-T^*
\end{array} \right)\!
\left( \begin{array}{c}
{\varphi^+} \\ \\ {\varphi^-}
\end{array} \right)
=E\! \left( \begin{array}{c}
{\varphi^+} \\  \\ {\varphi^-}
\end{array} \right)\! ,
\end{displaymath}
where we have defined  $T^*=(T_d+T_s)/2$, ${\Delta T}=T_d-T_s$.
A nice feature of this representation
is that for ${\Delta T}=0$ the equations for
$\varphi^+$ and $\varphi^-$ decouple, each becoming a simple harmonic
oscillator problem. Motivated by this
fact, we consider a variational solution
to the full problem  in terms of normalized ground state
harmonic oscillator wave-functions,
\begin{equation}
\varphi_k^\pm(x)= \sqrt{\sigma_\pm \over \sqrt{\pi}} e^{-\sigma_\pm^2
(x-x_k)^2/2}.
\label{varsol}
\end{equation}
\noindent The variational parameters $\sigma_+$ and $\sigma_-$  will
be determined by minimization of the eigenvalue $\langle E\rangle$.
If $\sigma_+=\sigma\cos\vartheta$ and $\sigma_-
=\sigma\sin\vartheta$, the resulting expression for $\langle
E\rangle$ can be minimized with respect to $\sigma^2$ analytically.
This leads to
\begin{eqnarray}
{\langle E\rangle \over{\Delta T}} = {T -
T^*\over{\Delta T}} + {1\over 4}\biggl({\hbar\omega_c\over{\Delta T}}\biggr)
\biggl[(1&+&{\epsilon_v})x +(1-{\epsilon_v}){1\over x}\biggr]
\nonumber \\
&-&{1\over 2}\sqrt{{2x\over 1+x^2}},
\label{Emin3}
\end{eqnarray}
where $x=\tan\vartheta$. The last equation must be minimized
numerically with respect to $x$, and is governed by the parameter
$\Lambda=\hbar\omega_c/{\Delta T}$. In the low field limit, set by
$\Lambda\ll 1$, $\sigma_+\approx\sigma_-\approx
\sqrt{m\omega_c/\hbar}$, while in the high field limit, $\Lambda\gg
1$, $\sigma_\pm\approx\sqrt{m\omega_c/\hbar}
[(1\pm{\epsilon_v})/(1\mp{\epsilon_v})]^{1/4}$. It follows that at least
intermediate values of $\Lambda$ are required for appreciable effects from
$s$-$d$ mixing to occur. Otherwise  the $s$ component
effectively
vanishes.

Solutions to Eq.(\ref{Emin3}) with $\langle E\rangle=0$ give the
dependence of the upper critical field $H_{c2}$ on the temperature.
Whenever a finite admixture of the $s$ component is present,
a characteristic upward curvature is found near the critical temperature in
$H_{c2}(T)$. Such curvature has been experimentally observed in both
La-Sr-Cu-O and Y-Ba-Cu-O compounds \cite{hidaka}, and has
also been interpreted as a result of $s$-$d$ mixing by
Joynt\cite{joynt}.

We next construct a vortex lattice. Consider a
periodic solution of the form

\begin{equation}
\chi_{d/s}({\bf r})=\sum_n c_n e^{inqy}[\varphi^+_n(x)\pm\varphi^-_n(x)],
\label{per}
\end{equation}
where $\varphi^\pm_k(x)$ are defined by (\ref{varsol}) and
$k=qn$, ($n$ integer), which gives periodicity in $y$  with
period $L_y=2\pi/q$. Solution (\ref{per}) will also  be periodic in
$x$ provided that the constants $c_n$ satisfy the condition
$c_{n+N}=c_n$ for some integer $N$. In what follows we
consider only the case of $N=2$ so that $c_{2n}=c_0$ and
$c_{2n+1}=c_1$. The period in the $x$ direction is $L_x=2l^2q$, and
it also follows that $BL_xL_y=2(hc/e^*)$ $=2\Phi_0$, i.e. there are
two flux quanta per unit cell.  The resulting lattice may be thought
of as centered rectangular with two quanta per unit cell or,
equivalently, as an oblique lattice with lattice vectors of equal
length and one flux quantum per unit cell. The parameter $q$
controls the shape of the vortex lattice, and it is customary to
describe this shape by the ratio, $R={L_x/ L_y} =(l^2/\pi) q^2$.
$R=1$ corresponds to the square, while $R=\sqrt{3}$ corresponds to
the triangular vortex lattice. The restriction to centered
rectangular lattices is made primarily for computational
convenience.  However it is compatible with recent experiments on
YBa$_2$Cu$_3$O$_7$ which show evidence for an oblique vortex
lattice with nearly equal lattice constants \cite{keimer1}.

At $B\approx H_{c2}(T)$ all solutions of the form (\ref{per}) are
degenerate. It is the fourth order terms in the free energy that lift this
degeneracy below $H_{c2}$ and determine the vortex lattice configuration.
Consider the solution to the full nonlinear problem of the form
$(s,d)=c(\chi_s,\chi_d)$ where $c$ is an arbitrary constant. Then the total
integrated free energy of the
system becomes
$F =|c|^2\langle f_2[\chi_s,\chi_d] \rangle + |c|^4\langle f_4[\chi_s,\chi_d]
\rangle$,
where $\langle \dots \rangle$ means integration over the volume of the system,
and $f_2$ and $f_4$ stand for quadratic and quartic parts of the free energy
density (\ref{fgl}) respectively.
One can now minimize the
total free energy
$F$ with respect to $|c|^2$ to obtain
\begin{equation}
F_{min}= -{\langle f_2 \rangle^2 \over 4\langle f_4 \rangle }
\equiv-E^2 \beta_A^{-1},
\label{fmin}
\end{equation}
where $\beta_A$ is the generalization of the usual Abrikosov parameter
\cite{abrikosov}.

We have studied the dependence of $\beta_A$ on $R$ in various
regions of the parameter space. The qualitative
features of this dependence are most strongly influenced by the
parameter $\epsilon_v$. In particular, for values  ${\epsilon_v}$ close to 0,
$\beta_A(R)$ is minimized by $R_{min}=\sqrt{3}$, i.e.
the triangular lattice is stabilized. This is to be expected, since
in this limit $\sigma_+\simeq\sigma_-$, so the $s$-component is
suppressed and the usual one component
solution\cite{abrikosov,kleiner} is found. The consistency of our solution is
confirmed in this limit where we recover the  correct value of the
Abrikosov ratio $\beta_A(\sqrt{3})=1.1596$ as quoted by Kleiner {\em
et al.}\cite{kleiner}. However as ${\epsilon_v}$ is increased, the
minimum of $\beta_A$ moves toward smaller values of
$R_{min}<\sqrt{3}$ signaling that an oblique vortex lattice is
preferred. Finally, at some value of ${\epsilon_v}$ (which depends on other
parameters) the minimum of $\beta_A$ reaches $R_{min}=1$,
characteristic of the square vortex lattice. Further increase of
${\epsilon_v}$ has no effect on the lattice structure which remains square.
The typical dependence of $\beta_A$ on $R$ is displayed in Figure
\ref{fig:beta}.

For all different parameter combinations studied, the
minimum $R_{min}$ varies continuously with ${\epsilon_v}$.  Varying other
parameters
changes somewhat the quantitative behavior of this dependence, but the
essential qualitative features described above remain intact.
For example, increasing the parameter $\beta_4$ makes the transition
from triangular to square lattice somewhat sharper (see inset of Figure
\ref{fig:beta}).  One would
also expect the inclusion of three dimensional screening effects (i.e.
replacing $B=H$ with $B=H+\langle h_s\rangle$, where
$\langle h_s \rangle$ is calculated self-consistently) to change the results
quantitatively, but not qualitatively.

In practice this continuous dependence would mean that in a
$d$-wave superconductor one would expect to observe a general
oblique vortex lattice, unless the material is in one of the
limiting regimes where ${\epsilon_v}$ is very small or very
large. Such an  oblique lattice structure has in fact been
recently observed by SANS on YBa$_2$Cu$_3$O$_7$ single crystals
in strong magnetic fields  parallel to the $c$ axis by Keimer
{\em et al.}\cite{keimer1}. These authors reported an oblique
lattice structure with nearly equal lattice constants and an
angle of $\phi=73^\circ$ between primitive vectors. Our
phenomenological theory is consistent with any angle $\phi$ in
the interval $(60^\circ,90^\circ)$, including that found
experimentally. We display an example of a general oblique vortex
lattice obtained by explicitly evaluating amplitudes of $s$ and
$d$ components of the order parameter from Eqs. (\ref{per}) in
Figure \ref{fig:dscont}. Comparison of Figures \ref{fig:dscont}a
and \ref{fig:dscont}b reveals that the non-trivial topological
structure of the $s$ component persists even in this high field
regime.

Keimer {\em et al.} further report that  one principal axis of
the oblique unit cell is always found to coincide with the (110)
or $(1\bar{1}0)$ direction of the YBa$_2$Cu$_3$O$_7$ crystal.
This is at variance with our results, since we find that one
lattice vector of the larger rectangular cell is oriented along
(100) or (010), even in the presence of a small orthorhombic
distortion. However, we note that the energy cost of rotating the
vortex lattice is small compared to the energy needed to deform
the lattice. It is thus possible that (110) twin boundaries,
where the order parameter is weakened, bind lines of vortices and
hence orient one of the oblique lattice vectors along (110) as is
found experimentally.

This work has been partially supported by the Natural Sciences and
Engineering Research Council of Canada, Ontario Centre for Materials
Reasearch, and by the National Science Foundation under Grant
No. DMR-91-20361. We would also like to thank the Aspen Center for
Physics, where this collaboration was initiated.



\begin{figure}
\caption[]{ Amplitude of the $s$-wave component
 along the $x$-axis (solid line) and along the diagonal $x=y$ (dotted line)
normalized to the bulk value $d_0$. The parameters used are:
$\gamma_s=\gamma_d=\gamma_v$, $\alpha_s=10|\alpha_d|$, $\beta_1=\beta_3=0$,
and $\beta_4=0.5\beta_2$.  The inset shows schematically the positions
of the $s$-wave vortices and their relative windings.}
\label{fig1}
\end{figure}

\begin{figure}
\caption[]{ Abrikosov ratio $\beta_A$ as a function of the lattice geometry
factor $R=L_x/L_y$ for different values of ${\epsilon_v}$ and
$T_s=0.5 T_d$, $T=0.75 T_d$, $\beta_1=\beta_2=\beta_3=\beta_4=
1$, and $B=0.8 H_{c2}$. The inset shows the dependence of the minimum
$R_{min}$ on the parameter ${\epsilon_v}$ for different values of $\beta_4$.}
\label{fig:beta}
\end{figure}

\begin{figure}
\caption[]{Contour plot of the amplitudes of (a) $d$  component and (b) $s$
component of the order parameter. GL parameters are the same as in Figure
\ref{fig:beta} with ${\epsilon_v}=0.45$, resulting in an oblique vortex
lattice with $R_{min}=1.29$ and the angle between primitive vectors
$\phi=76^\circ$.  The lightest regions correspond to the largest
amplitudes.}
\label{fig:dscont}
\end{figure}

\end{document}